\documentclass[conference]{IEEEtran}
\usepackage{amsmath}
\usepackage{amsfonts}
\usepackage{amstext}
\usepackage{epsfig}
\usepackage{amssymb}
\usepackage{cite}
\usepackage{hhline}
\usepackage{multirow}
\usepackage{framed}
\usepackage{xcolor}
\usepackage{lipsum}
\usepackage{enumerate}
\usepackage{color}
\usepackage{caption}
\usepackage{subcaption}
\usepackage{graphicx}
\usepackage{algorithm}
\usepackage{algpseudocode}
\usepackage{mathtools}
\usepackage{booktabs}
\usepackage{bm}
\usepackage{cite}
\usepackage{url}
\usepackage{makecell}
\usepackage[english]{babel}
\usepackage{graphicx}
\usepackage[normalem]{ulem}
\usepackage{enumitem}
\usepackage{cite}

\usepackage{pgfplots}
\pgfplotsset{compat=newest}
\usepackage{tikz}
\usetikzlibrary{plotmarks}
\usetikzlibrary{arrows.meta}
\usepgfplotslibrary{patchplots}
\usepackage{grffile}

\usetikzlibrary{shapes,arrows}

\usetikzlibrary{intersections}

\ifCLASSINFOpdf
\def\BibTeX{{\rm B\kern-.05em{\sc i\kern-.025em b}\kern-.08em
		T\kern-.1667em\lower.7ex\hbox{E}\kern-.125emX}}
\begin{document}

\title{Learning-based Load Balancing Handover in Mobile Millimeter Wave Networks}


\author{\IEEEauthorblockN{Sara Khosravi, Hossein S. Ghadikolaei, and Marina Petrova}
	\IEEEauthorblockA{School of EECS, KTH Royal Institute of Technology, Stockholm, Sweden \\
		Emails:\{sarakhos, hshokri, petrovam\}@kth.se}
}

\maketitle
\renewcommand{\figurename}{Fig.}
\newcommand{\argmax}{\mathrm{argmax}}
\newcommand{\Tx}{\mathrm{Tx}}
\newcommand{\Rx}{\mathrm{Rx}}
\newcommand{\Base}{\mathrm{Base}}
\newcommand{\maximize}{\mathrm{maximize}}
\newcommand{\SNR}{\mathrm{SNR}}
\newcommand{\R}{\mathrm{R}}
\newcommand{\PS}{\mathrm{PS}}
\newcommand{\LoS}{\mathrm{LoS}}
\newcommand{\NLoS}{\mathrm{NLoS}}
\newcommand{\C}{\mathrm{C}}
\newcommand{\Rate}{\mathrm{Rate}}
\newcommand{\BS}{\mathrm{BS}}
\newcommand{\UE}{\mathrm{UE}}
\newcommand{\T}{\mathrm{T}}
\newcommand{\Aging}{\mathrm{Aging}}
\newcommand{\D}{\mathrm{D}}
\newcommand{\DM}{\mathrm{DM}}
\newcommand{\HO}{\mathrm{HO}}
\newcommand{\PL}{\mathrm{PL}}
\def\TT{\mathsf{T}}
\def\HH{\mathsf{H}}
\newcommand{\Pilot}{\mathrm{Pilot}}
\newcommand{\tr}[1]{\mathrm{tr}\left({#1}\right)}

\allowdisplaybreaks

\begin{abstract}
Millimeter-wave (mmWave) communication is a promising solution to the high data rate demands in the upcoming 5G and beyond communication networks. When it comes to supporting seamless connectivity in mobile scenarios, resource and handover management are two of the main challenges in mmWave networks. In this paper,
we address these two problems jointly and propose a learning-based load balancing
handover in multi-user mobile mmWave networks.
Our handover algorithm selects a backup base station and allocates the resource to maximize the sum rate of all the users while ensuring a target rate threshold and preventing excessive handovers. We model the user association as a non-convex optimization problem. Then, by applying a deep deterministic policy gradient (DDPG) method, we approximate the solution of the optimization problem. Through simulations, we show that our proposed algorithm minimizes the number of the events where a user's rate is less than its minimum rate requirement and minimizes the number of handovers while increasing the sum rate of all users.
\end{abstract}


\section{Introduction}
The expected growth of mobile traffic and high data rate demands has triggered the design of communication networks over millimeter-wave (mmWave) bands \cite{rappaport2013broadband}. MmWave networks promise Gbps data rates over directional links. However, high penetration loss and vulnerability to rapidly varying channel conditions and blockages are the principal challenges to overcome in mmWave bands \cite{andrews2016modeling}.
In order to provide proper coverage and capacity, dense base station (BS) deployments are necessary, and frequent handovers may occur in scenarios where both the user and the obstacles are mobile. The latter increases the signaling overhead and the service delay due to more frequent channel estimation and cell association.
Thus, handover and resource allocation during user association is an important  issue in mmWave networks \cite{ye2013user}.

The problem of user association to achieve load balancing in mmWave networks has received considerable attention in recent years. The authors in \cite{ye2013user} proposed a distributed algorithm via dual decomposition to achieve load balancing among different tiers in the heterogeneous network. The authors in \cite{xu2018adaptive, alizadeh2019load } formulated the association as a mixed-integer optimization and proposed a heuristic approach and a polynomial-time algorithm to solve it. However, the sensitivity of the association solution to the network dynamics \cite{shokri2015millimeter} causes the re-execution of the complicated association problem.

In \cite{MDP}, the association problem is modeled as a Markov decision process framework.
The authors in \cite{khan2019reinforcement} leverages a distributed deep reinforcement learning algorithm to find the optimal solution of the user-association problem in V2X communication networks.
The authors in \cite{sun2017reinforcement} proposed a reinforcement learning handoff policy to reduce the number of handoffs in heterogeneous networks. However, the long-term effect of handovers in order to maximize the sum rate of all the UEs along trajectories and at the same time reduce the number of the handovers and ensure the rate requirement per UE to provide a reliable connection, have not been addressed in the aforementioned works.

In this paper, we propose a learning-based load balancing handover in a multi-user mobile mmWave networks. Our algorithm aims to associate the UEs and allocate the required resources to maximize the sum rate of all UEs moving along different trajectories in the environment
and at the same time minimize the number of  handovers and the probability of the events in which each UE's average rate is less than a pre-defined data rate threshold. We model the user association problem as a non-convex optimization problem 
and deploy a machine-learning tool, namely deep deterministic gradient method, to approximate the solution of this optimization problem. Leveraging machine-learning in mobile mmWave, as the main decision-making tool, is motivated by the fact that it can learn the geometry of the environment based on the statistic data and past experiences. In our proposed approach, most of the computational tasks are done during offline training to approximate the solution of the optimization problem based on  databases of the dynamic environment from the past experiences. Then, the optimal policy of the offline training process is validated during the online test. The simulation results show that our proposed load balancing handover method increases the sum rate of the UEs, reduces the number of the outage events (when the average achieved rate becomes less than the rate requirement) and reduces the number of handovers in comparison with two benchmarks, namely random backup benchmark and the worst connection swapping (WCS) algorithm (similar idea to \cite{alizadeh2019load}).

The rest of the paper is organized as follows. In Section \ref{system model}, we introduce our system model and explain our problem formulation. In Section \ref{method}, we introduce our learning-based load balancing handover method. We present the numerical results in Section \ref{results}, and conclude our work in Section \ref{conclusions}.

\textit{Notations:} 
Throughout this paper, matrices, vectors and scalars are represented by bold upper-case ($\mathbf{X}$), bold lower-case ($\mathbf{x}$) and non-bold ($x$) letters, respectively. The transpose, the conjugate transpose and the $\ell_{2}$-norm of a vector $\mathbf{x}$ denote $\mathbf{x}^\T$, $\mathbf{x}^\HH$ and $\|\mathbf{x}\|$, respectively. We define set $[L]=\{1,2,..,L\}$ for any integer $L$. The indicator function $1\{\cdot\}$ is equal to one if the constraint inside $\{\cdot\}$ is satisfied.

\section{System Model and Problem Formulation}
\label{system model}
We consider a downlink mmWave network with $| \mathcal{B}|$ BSs and $ |\mathcal{U}|$ mobile user equipments (UEs), where each UE is served by only one BS.
We assume each UE is moving through a specific trajectory with length $L \in \mathcal{N}$ where $\ell(i)$ is the location of UE $i \in \mathcal{U}$. Each BS is equipped with $N_{\BS}$ antennas, and each UE is equipped with $N_{\UE}$ antennas.
We assume that all BSs have full buffers and fixed transmit powers. Next, we explain the channel model and problem formulation. 

\subsection{Channel Model}
\label{model_A}
Considering a narrow band cluster 3D channel model \cite{andrews2016modeling}, the channel matrix $\mathbf{H}\in\mathbb{C}^{N_{\BS}\times N_{\UE}}$ between UE $i$ in location $\ell(i)$ of a trajectory (with $L$ points) and BS $j\in \mathcal{B}$ is fixed during a coherence interval (CI) and can be defined as:
\begin{equation}
	\mathbf{H}(i,j)
	\!=\!{\frac{1}{\sqrt{R}}}\sum_{c=1}^{C}\sum_{r=1}^{R} h_{r,c} \mathbf{u}_{\text{\tiny UE}}(\theta^{\UE}_{r,c},\phi^{\UE}_{r,c}) \mathbf{u}^\HH_{\text{\tiny BS}}(\theta^{\BS}_{r,c},\phi^{\BS}_{r,c}),
\label{H}
\end{equation}
where $C$ is the number of path clusters and $R$ is the number of subpaths in each cluster. Each subpath have horizontal and vertical angle of arrivals (AoAs), $\theta^{\UE}_{r,c},\phi^{\UE}_{r,c}$, and horizontal and vertical angle of departures (AoDs), $\theta^{\BS}_{r,c},\phi^{\BS}_{r,c}$. The complex gain of $r$-th subpath of cluster $c$ is $h_{rc}$ which includes both the path loss and small scale fading \cite{andrews2016modeling}. These parameters are generated based on different distribution as given in \cite[Table I]{andrews2016modeling}. For the sake of notation simplicity, we drop the notation $i$ and $j$ from the the channel parameters, whenever they are clear from the context. $\mathbf{u}_{s}(.)\in \mathbb{C}^{N_{s}}, s\in\{\UE,\BS\} $ is the vector response function of the BS and UE antenna arrays to the AoAs and AoDs. In this work, we consider a half wavelength uniform planar arrays of antennas both at the BS and the UE sides which can be defined as \cite{hemadeh2017millimeter}:
\begin{equation}
 \mathbf{u}_{s}(\theta^s,\phi^s)=[1,..., e^{j\pi[n_\BS \sin(\theta)\cos (\phi)+n_\UE \sin(\theta)\sin(\phi)]},... ]^T
\end{equation}
where $1\leq n_\BS\leq N_\BS-1$, $1\leq n_\UE\leq  N_\UE-1$, and $s\in\{\UE,\BS\}$.

For the blockage model, we use the probability functions obtained based on the New York City measurements in \cite{samimi2015probabilistic} to define the probability of LoS and NLoS states of each link:
\begin{subequations}
\begin{align}
&p_{\LoS}(d)=\left[\min \left(\frac{27}{d},1\right).\left(1-e^{-\frac{d}{71}}\right)+e^{-\frac{d}{71}}\right]^2
\\& p_{\NLoS}(d)=1-p_{\LoS}(d),
\end{align}	
\end{subequations}
where $d$ is the 3D distance between UE and BS in meters. We model the pathloss for LoS and NLoS links as:
\begin{equation}
\PL(d)[\text{dB}]=10 \log_{10}\left(\frac{4\pi d_{0}}{\lambda}\right)^2+10 \hat{n}  \log_{10}\left(\frac{d}{d_{0}}\right)+X_{\mu},
\end{equation}
where $d_0$ is the close-in free space reference distance which in this work $d_0=1$, $\lambda$ is the wavelength and $\hat{n} $ is the path loss exponent which has different amount depending on the LoS or NLoS links. $X_{\mu}$ is a zero mean Gaussian random variable with a standard deviation $\mu$ in dB which represents the shadow fading for LoS or NLoS links. These parameters are given in \cite [Table V and VI]{rappaport2015wideband}.

Under the assumption of using capacity achieving codes, the achievable rate, between BS $j$ and UE $i$, can be approximated with link capacity $c(i,j)$ which is defined as:
\begin{equation} \label{eqn:c(i,j)}
	c(i,j)
	= W\log_{2}\left(1
	+\frac{p \lvert\mathbf{w}(i)^{\HH}\mathbf{H}(i,j) \mathbf{f}(i,j)\rvert^2}{\sigma^{2}W+I(i,-j)}\right),
\end{equation}
\begin{equation*}
I(i,-j)=\sum_{j'\in \mathcal{B}, j'\neq j}\sum_{i'}p\lvert\mathbf{w}(i)^{\HH}\mathbf{H}(i,j')\mathbf{f}(i',j')\rvert^2 ,
\end{equation*}
where $p$ is the transmit power, $W$ is the system bandwidth, $\sigma^{2}$ is the noise power level, $\mathbf{f}\in \mathbb{C}^{N_\BS}$ is the beamforming vector in the BS side and $\mathbf{w}\in \mathbb{C}^{N_\UE}$ is the combining vector in the UE side. The combing and beamforming vectors are found from the left and right singular vectors of $\mathbf{H}$. The interference, $I(i,-j)$ is coming from other BSs ($j'\neq j$) which are sending signals to their active UEs $i'$. In this work we omit the interference effect of other active UEs of the BS $j$ and leave it as a future work. 

We define a zone based on a certain geographical area with a specific number of BSs. We consider one agent as the decision-maker for each zone that connects to all the BSs in its zone.  


\subsection{Problem Formulation}
We consider the problem in the discrete-time domain, where the duration of each time slot is equal to the duration of the CI.
As it is shown in Fig. \ref{fig:fig1}, each time slot $t$ starts with data transmission within a time of less than one CI. Afterward, we assume that the current CI ends, and the next CI begins while the location dynamics follow a mobility pattern. The new CI ends after the next data transmission.
At the end of the time slot $t$, channel estimation, resource allocation, and handover decision making in the new location is executed, as shown in Fig. \ref{fig:fig1}. We assume the channel estimation during mini-slot $S$ and mini-slot $B$ are done based on an efficient beamforming algorithm in \cite{sara} which has low signaling overhead during the beamforming phase.

For each time slot $t$, we define variables $x^t(i,j)\in[0,1]$ for all $i\in\mathcal{U}, j\in\mathcal{B}$, where it represents the normalized shared resources of BS $j$ for UE $i$.
As a result, $x^t(i,j)>0$ if and only if BS $j$ is serving UE $i$.
Therefore, in time slot $t$, the achieved rate of UE $i$ is
\begin{equation*}
\R^t(i) = x^t(i, j_S^t(i))c^t(i, j_S^t(i))
= \sum_{j\in\mathcal{B}}{x^t(i,j) c^t(i,j)},
\end{equation*}
where $c^t(i,j)$ is the capacity defined in \eqref{eqn:c(i,j)}, and $j_S^t(i)$ is the index of the serving BS of UE $i$.

\tikzset{%
	bodyy/.style={inner sep=0pt,outer sep=0pt,shape=rectangle,draw,thick},
	dimen/.style={<->,>=latex,thin,every rectangle node/.style={fill=white,midway,font=\sffamily}},
	dimen1/.style={->,>=latex,thin,every rectangle node/.style={fill=white,midway,font=\sffamily}},
	symmetry/.style={dashed,thin},
}
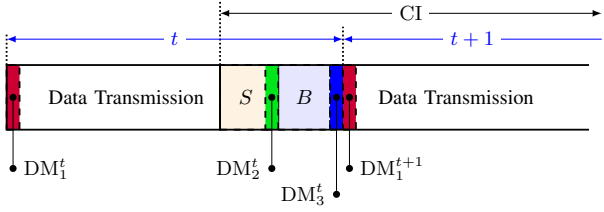
\begin{figure}[t]
	\centering
	\scalebox{0.86}{\hspace{-3mm}\small{\begin{tikzpicture}


    \node [bodyy,dashed,minimum height=1cm,minimum width=8mm,anchor=south west,fill=yellow!50!red!10] (S) at (3.3,0) {$S$};
    \node [bodyy,dashed,minimum height=1cm,minimum width=8mm,anchor=south west, fill=blue!10!white!] (B) at (4.2,0) {$B$};
    \node [bodyy,dashed,minimum height=1cm,minimum width=2mm,anchor=south west, fill=blue!10!green!] (C2) at (4,0) {};
        \node [bodyy,dashed,minimum height=1cm,minimum width=2mm,anchor=south west, fill=blue] (C1) at (5,0) {};
       \node [bodyy,dashed,minimum height=1cm,minimum width=2mm,anchor=south west, fill=blue!20!red!] (C4) at (5.2,0) {};
    \node [bodyy,thick, minimum height=1cm,minimum width=6cm,anchor=south west] (body1) at (3.3,0) {};
  \node [bodyy,draw=none,minimum height=12mm,minimum width=3.5mm,anchor=south west,fill=white]
    (body3) at (9,-0.1) {};

\node [bodyy,dashed,minimum height=10mm,minimum width=2mm,anchor=south west,fill=blue!20!red!](body3) at (0,0) {};
\node [bodyy,thick,minimum height=1cm,minimum width=3.3cm,anchor=south west] (body2) at (0,0) {};
\node [bodyy, draw=none, minimum height=1cm,minimum width=3.5cm,anchor=south west] at (5.2,0) {Data Transmission};
\node [bodyy, draw=none, minimum height=1cm,minimum width=4.5cm,anchor=south west] at (-0.4,0) {Data Transmission};
\draw[densely dotted,thick] (0,1) -- ++(0,+0.4) coordinate (D1) -- +(0,+3pt);
\draw[densely dotted,thick] (3.3,1) -- ++(0,+0.8) coordinate (D11) -- +(0,6pt);
\draw[densely dotted,thick] (5.2,1) -- ++(0,0.4) coordinate (D2) -- +(0,5.5pt);
\draw [dimen,blue] (D1) -- (D2) node {$t$};
\draw[draw=none] (9.2,1) -- ++(0,0.4) coordinate (D3) -- +(0,5.5pt);
\draw[draw=none] (9.2,1) -- ++(0,0.8) coordinate (D33) -- +(0,5.5pt);
\draw [dimen1,blue] (D3) -- (D2) node {$t+1$};
\draw [dimen] (D11) -- (D33) node {$\mathrm{CI}$};
\node[outer sep=0pt,circle, fill,inner sep=1.1pt] (C11) at (4.1,0.5) {};
\node[outer sep=0pt,circle, fill,inner sep=1.1pt, label={left:$\DM^t_{2}$}] (C12) at (4.1,-0.6) {};
\draw (C11) -- (C12);
\node[outer sep=0pt,circle, fill,inner sep=1.1pt] (C21) at (5.1,0.5) {};
\node[outer sep=0pt,circle, fill,inner sep=1.1pt, label={left:$\DM^{t}_{3}$}] (C22) at (5.1,-1) {};

\draw (C21) -- (C22);
\node[outer sep=0pt,circle, fill,inner sep=1.1pt] (C31) at (0.1,0.5) {};
\node[outer sep=0pt,circle, fill,inner sep=1.1pt, label={right:$\DM^{t}_{1}$}] (C33) at (0.1,-0.6) {};
\draw (C31) -- (C33);

\node[outer sep=0pt,circle, fill,inner sep=1.1pt] (C41) at (5.3,0.5) {};
\node[outer sep=0pt,circle, fill,inner sep=1.1pt, label={right:$\DM^{t+1}_{1}$}] (C44) at (5.3,-0.6) {};
\draw (C41) -- (C44);
\end{tikzpicture} }}
	\caption{Time segmentation. Mini-slot $S$ and mini-slot $B$ are related to the channel estimation toward serving and backup BSs, respectively. $\DM^t_{1}$ is the decision making phase regarding the choice of backup $\BS$ during mini-slot $B$. $\DM^t_{2}$ is the decision making phase regarding the handover execution, and $\DM^t_{3}$ is related to the resource allocation phase.}
	\label{fig:fig1}
\end{figure}

In this paper, we consider multiple mobile UEs moving through different trajectories.
The goal is to maximize the sum of achievable rates of the UEs, while minimizing the number of handovers and the number of UEs which have the rates less than their rate requirement. For the latter, we define a window of recent achieved rates and ensure that the UEs averaged rate over this window is larger than an $\R_{\rm th}$. To this end, we define functions $F_1$, $F_2$ and $F_3$ as follows.
\begin{itemize}[leftmargin=*]
\item $F_1$ is the averaged trajectory rate of the UEs, defined as the aggregated rates of the UEs along their own trajectories. Formally,
\begin{IEEEeqnarray*}{rCl}
F_1
&=&\sum_{t=1}^T\sum_{i\in\mathcal{U}} \mathbb{E}\left[\R^t(i)\right]=\sum_{t=1}^T\sum_{i\in\mathcal{U}}\sum_{j\in\mathcal{B}}\mathbb{E}[ x^t(i,j) c^t(i,j)],
\end{IEEEeqnarray*}
where $T$ is the number of time slots during a trajectory, and the expectation is with respect to the randomness of $c^t(i,j)$.

\item $F_2$ is the expected number of UEs whose rate requirement has not been met. We define the rate requirement as the average rate of each UE in the last $K$ CIs, so
\begin{IEEEeqnarray*}{l}
F_2
= \mathbb{E}\left[\sum_{t=1}^T \sum_{i\in \mathcal{U}}1\left\{\frac{1}{K} \sum_{k=0}^{K-1} \R^{t-k}(i)\leq \R_{\rm th}(i) \right\}\right] \\
= \sum_{t=1}^T \sum_{i\in \mathcal{U}} \Pr\Big\{
\frac{1}{K} \sum_{k=0}^{K-1}\sum_{j\in\mathcal{B}} x^{t-k}(i,j)c^{t-k}(i,j) \leq \R_{\rm th}(i) \Big\},
\end{IEEEeqnarray*}
where  $\R_{\rm th}(i)$ is the minimum rate requirement of the UE $i$. Note that if $t-k \leq 0$, we assume $\R^{t-k}=0$.

\item $F_3$ is the expected value of the number of handovers. The number of handovers is
$1\left\{ \sum_{j\in\mathcal{B}} x^t(i,j) x^{t-1}(i,j) = 0 \right\}$ for UE $i$ in time slot $t$.
Hence, $F_3$ is
\begin{IEEEeqnarray*}{rCl}
F_3
&=& \mathbb{E}\left[\sum_{t=1}^T \sum_{i\in \mathcal{U}}
1\left\{\sum_{j\in\mathcal{B}} x^t(i,j) x^{t-1}(i,j) = 0 \right\}\right] \\
&=& \sum_{t=1}^T \sum_{i\in \mathcal{U}}
\Pr\left\{\sum_{j\in\mathcal{B}} x^t(i,j) x^{t-1}(i,j) = 0 \right\}.
\end{IEEEeqnarray*}
\end{itemize}

Therefore, we propose the optimization problem as follows
\begin{subequations}
\begin{alignat}{3}
&\max_{\substack{\{x^t(i,j)\}_{t,i,j} 
}}
&\quad& F_1- \lambda_t F_2 - \lambda_h F_3 \label{eq:optProb} \\
&\mathrm{subject~to}
&& \sum_{j\in \mathcal{B}}1\{x^t(i,j)> 0\}\leq 1, \forall i\in \mathcal{U}, t\in[T] \label{eq:constraint1}\\
&&& \sum_{i\in \mathcal{U}}x^t(i,j)\leq 1,\quad\forall j\in\mathcal{B}, t\in[T] \label{eq:constraint2}\\
&&& x^t(i,j)\geq 0, \quad\forall i\in \mathcal{U}, j\in \mathcal{B}, t\in[T], \label{eq:constraint3}
\end{alignat} \label{optimization}
\end{subequations}
\!\!\!where $\lambda_t$ and $\lambda_h$ are positive constants controlling the importance of $F_2$ and $F_3$, respectively.
We assume that all variables $\lambda_t$, $\lambda_h$ and $\lambda_t/\lambda_h$ are large enough.
Thus, reducing the probability of UE rate becomes less than threshold is more important than reducing the number of handovers, and reducing the number of handovers is more important than increasing the total sum rate of UEs.
Constraint \eqref{eq:constraint1} guaranties that each UE is served by atmost one BS.
Constraint \eqref{eq:constraint2} refers to the resource limitation of BSs.
Therefore, each BS allocates a ratio of its resource to UEs, and the sum of ratios cannot be more than $1$. According to \eqref{eq:constraint3}, the amount of resource allocation is a positive real number.

The optimization problem \eqref{optimization} is nonlinear because of \eqref{eq:constraint1}; as a result, the computational complexity of solving this problem is high. Further, the rate allocation policy must be causal, i.e., the optimal $x^t(i,j)$ is independent of $c^{t+1}(\cdot , \cdot), \ldots, c^T(\cdot , \cdot)$ given $c^1(\cdot , \cdot), \ldots, c^t(\cdot , \cdot)$, for all $i\in\mathcal{U},j\in\mathcal{B},t\in[T]$. 
Therefore, the impact of the selection of the backup BSs propagates in time and may affect the UEs' QoS in later times.
Hence, it is crucial to consider the long-term benefits of selecting backup BSs besides their temporal effects on UEs' QoS.
Moreover, in order to select backup BSs, one needs to predict and model the UEs' QoS, in future time slots, which may not be affordable due to the high mobility of UEs and obstacles in mmWave networks.
These requirements motivate us to transfer the optimization problem \eqref{optimization} to the reinforcement learning (RL) problem.

\section{Reinforcement Learning-based Handover}
\label{method}
In this work, we propose to keep a set of backup BSs for each set of UEs, which may need to do a handover in the next time slot.
In the RL framework, the objective function is transformed into the reward function, and the network constraints are transformed into the feasible state and action spaces.
In the following we will describe the Markov decision process formulation of the problem.
Then, we will explain our proposed handover process.

The optimal solution ${x^t}^*(i,j)$, in \eqref{optimization}, can be found in two steps: finding the non-zero elements of ${x^t}^*(i,j)$, and the optimal value of ${x^t}^*(i,j)$.
It is the same as first setting $j_S^t(i)\in\mathcal{B}$ as the serving BS of UE $i$ in time slot $t$, and then finding $\alpha^t(i)$, which is the ratio of resources of BS $j_S^t(i)\in\mathcal{B}$ allocated to UE $i$.
As a result, we can write the optimization as
\begin{IEEEeqnarray*}{l}
\max_{\substack{\{x^t(i,j)\}_{t,i,j}}}
\!\!\!\!\!\!F_1 - \lambda_t F_2 - \lambda_h F_3
=\!\! \max_{\{j_S^t(i)\}_{t,i}}
\!\!G\left(j_S^t(1),\ldots,j_S^t(\lvert\mathcal{U}\rvert\right),
\end{IEEEeqnarray*}
where
\begin{equation*}
G\left(j_S^t(1),\ldots,j_S^t(\lvert\mathcal{U}\rvert\right)
= \max_{\substack{\{\alpha^t(i)\in[0,1]\} }_{t,i}}
F_1 - \lambda_t F_2 - \lambda_h F_3
\end{equation*}
In order to reduce the complexity, we evaluate suboptimal ${\hat j^t}_S(i)$ utilizing RL framework, then evaluating $\alpha^t(i)$ based on a fixed policy.

\subsection{Markov Decision Process Formulation}
RL problems are formulated utilizing the idea of the Markov decision process (MDP), which is based on the agent's interaction with different states of the environment to maximize the expected long-term reward. The agent is the main decision-maker who can sit on the edge cloud.
Here, we define the elements of an MDP.

\subsubsection{State Space}
The sate space describes the environment in which the agent is interacting by selecting different actions and observing the reward.
We define the state at time slot $t$ in multi-user mmWave network as:

\begin{IEEEeqnarray*}{l}
s^{t} = \\
(\ell^{t}(i), j_{S}^{t}(i), c^{t}(i, j^{t}_{S}(i)), \alpha^{t}(i),\R^{t-1}(i), \ldots , \R^{t-K+2}(i))_{i\in\mathcal{U}},
\end{IEEEeqnarray*}
where $\ell^{t}(i)$ denotes the location of UE $i$;
$j_{S}^{t}(i)\in\mathcal{B}$ is the serving BS of UE $i$;
$c^{t}(i, j^{t}_{S}(i))$ is the capacity of the channel between UE $i$ and BS $j_{S}^{t}(i)$.
Note that $c^{t}(i, j^{t}_{S}(i))$ is a random value due to the randomness of the channel gain (fading and blockage).
$\alpha^{t}(i)\in[0,1]$ is the proportion of the resources of BS $j_{S}^{t}(i)$ allocated to UE $i$.
All the variables mentioned earlier are evaluated at the start of the time slot $t$.
$\R^{t-1}(i), \ldots, \R^{t-K+2}(i)$ are the last $K-2$ observation of achieved rate of UE $i$ in the previous time slots.
The agent, at the beginning of each time slot, sees the input state $s^t$ and accordingly choose an action.

\subsubsection{Action Space}
In the next state, a set $\mathcal{U}_h \subseteq \mathcal{U}$ of UEs does the handover.
Hence, the action of the agent is to determine a backup for each one of the UEs $i\in\mathcal{U}_h$ for all $2^{\lvert\mathcal{U}\rvert}$ possible subsets $\mathcal{U}_h\subseteq\mathcal{U}$.
In other words, actions are functions
\begin{equation*}
A \colon \mathcal{P}(\mathcal{U}) \to \mathcal{B}^{\lvert\mathcal{U}\rvert}, \quad A(\mathcal{U}_h)=\mathbf{a},
\end{equation*}
from $\mathcal{P}(\mathcal{U})$, the set of all subsets of $\mathcal{U}$, to the set of all possible allocations among UEs and BSs.
Each allocation between UEs and BSs is denoted by a vector $\mathbf{a}\in\mathcal{B}^{\lvert\mathcal{U}\rvert}$ where $a_i$ denotes the possible serving BS for UE $i$ in the next time slot.
We suppose that if $i\notin\mathcal{U}_h$, then its serving BS is not changed in the next time slot.
Note that actions are the function of the state in the corresponding time slot. Hence, actions are determined at the beginning of time slots, during $\mathrm{DM}_1^t$ (see Fig.~\ref{fig:fig1}).

\subsubsection{Transition to the next state}
For each UE $i$, transition to the next location $\ell^{t+1}(i)$ is realized based on the current location $\ell^t(i)$.
Taking into account that one can predict the mobility of the UEs in mmWave reliably \cite{mobility1}, we assume the mobility model, including UEs trajectories, are available; as a result, the transition between locations are deterministic.
The UEs which require to do a handover are the ones with average rate of the last $K-2$ time slots, the current time slot, and the next time slot if transmission continues with the same serving BS and resource allocation, becoming less than the threshold:
\begin{IEEEeqnarray}{l}
\frac{1}{K}\left[
\sum_{\tau=t-K+2}^{t}c^\tau(i,{j}^{\tau}_{S}(i)) \alpha^\tau(i)
+c^{t+1}(i,{j}^{t}_{S}(i)) \alpha^t(i)\right] \nonumber\\
\qquad < \R_{\mathrm{th}}(i)
\Leftrightarrow i\in \mathcal{U}_h, \label{eqn:handover}
\end{IEEEeqnarray}
where the set $\mathcal{U}_h$ represents the UEs are required to do the handover.
Note that the estimation of the channel in the new CI is done in the mini time slot $S$ after data transmission (see Fig. \ref{fig:fig1}).
The handover procedure is explained in Section \ref{sec:handover}.
After the handover, the new serving BS and the proportion of resources allocated to each UE, in time slot $t+1$, are determined.

\subsubsection{Reward Function}
In finite-horizon MDP, total reward $r$ is the summation of per-time rewards, $r^t$, which are evaluated at the end of each time slot based on the corresponding state $s^t$, action $A^t$ and the next state $s^{t+1}$.
In our problem, per-time reward in time slot $t$ is the summation of the rates of UEs in time slot $t$ minus the loss for the handovers and average rates getting less than the thresholds.
Hence, we have
\begin{equation*}
r = \sum_{t=1}^T{\mathbb{E}\left[r^t(S^t, A^t, S^{t+1})\right]},
\end{equation*}
where,
\begin{IEEEeqnarray}{l}
r^t(s^t, A^t, s^{t+1}) \nonumber\\
\qquad= \sum_{i\in \mathcal{U}} \R^t(i)
- \lambda_t \sum_{i\in \mathcal{U}} 1\left\{\frac{1}{K} \sum_{k=-1}^{K-2} \R^{t-k}(i)\leq \R_{\rm th}(i) \right\} \nonumber\\
\qquad\quad- \lambda_h \sum_{i\in\mathcal{U}} 1\left\{j_S^t(i) \neq j_S^{t+1}(i) \right\}, \label{eqn:rt}
\end{IEEEeqnarray}
where
\begin{equation} \label{eqn:RtiMDP}
\R^t(i) = c(i,j_S^t(i)) \alpha^t(i).
\end{equation}

In order to find the optimal policy, due to the continuous and large number of the state space, one solution is Deep Q Network (DQN) algorithm.
In this algorithm, the action-value function is estimated based on the deep neural network function approximators.
However, DQN cannot straightforwardly be applied to the continuous or very large action space\cite{lillicrap2015continuous}.
Due to the large number of the action space in our algorithm, which is $\prod_{i=0}^{\lvert\mathcal{U}\rvert} \lvert\mathcal{B}\rvert^{i \binom{|\mathcal{U}|}{i}} = \lvert\mathcal{B}\rvert^{\lvert\mathcal{U}\rvert 2^{\lvert\mathcal{U}\rvert - 1}}$ (which is a large number even in a sparse network), we utilize deep deterministic policy gradient (DDPG) method \cite{lillicrap2015continuous} which is compatible with the large number of states and action spaces.

\subsection{DDPG Algorithm}
The DDPG is an off-policy, model-free, and online RL method, which has two main components: actor and critic. The actor, $\mu(s)$, takes the state $s^t$ as the input and returns the corresponding action $a^t$, which maximizes the long-term reward.
The critic, $Q(s^t,a^t)$, returns the expected long-term reward based on the state and action inputs.
It can be defined for discounted infinite horizon MDP as
\begin{equation} \label{Q}
Q(s^t,a^t)
= \mathbb{E}[r(s^t,a^t,s^{t+1})+\gamma Q(s^{t+1},\mu(s^{t+1}))]
\end{equation}
where $\gamma$ is the discount factor \footnote{The best choice of the discount factor in finite horizon MDP is equal $1$.  However, the DDPG algorithm has been designed for the infinite horizon and converges with discount factor less than $1$. Hence, in our algorithm, we choose the discount factor close to $1$.} of MDP reward, and
\begin{equation*}
\mu(s^t)
= \argmax_{a}Q(s^t,a^t).
\end{equation*}


In DDPG, it is assumed that $Q(s, a)$ is the scaler output of a neural network with inputs $s,a$,
with parameters $\theta^Q$.
Actor is also the output of a neural network with input $s$ and output $a$, where the parameters are represented by $\theta^\mu$.
In order to explore new states, a noise $\mathcal{N}$ is added to the output of the actor to get new random actions. After getting the initial state of a episode, for each time step $t$ the action input of critic neural network will be selected based on the current policy and the action exploration noise. All the transitions $(s^t,a^t,r^t,s^{t+1})$ are stored in the reply buffer. Then, the critic and the actor policy and their parameters  $\theta^Q$ and $\theta^\mu$, respectively, are updated by sampling a mini-batch with size $M$ from the reply buffer. We refer reader to \cite{lillicrap2015continuous} for more details on DDPG algorithm.

Note that in our algorithm, DDPG method is applied to the offline training phase to estimate the optimal policy. Then, the obtained policy is used to the online phase. The DDPG method can be also used for the online phase in order to estimate the optimal policy asymptotically. However, due to the computational complexity of the online training phase, it needs a simpler neural network model. For instance, it can apply a neural network with lower number of the hidden layers or it can explore most of the time to find a policy close to the optimal one and does not exploit.

\subsection{Handover Algorithm} \label{sec:handover}
The handover process starts after mini-slot $S$ (see Fig.~\ref{fig:fig1}). At $\mathrm{DM}_2^t$ backup BSs are chosen based on $A^t(\mathcal{U}_h)$ and we define $j_B^t(i)$ as the $i$-th element of $A^t(\mathcal{U}_h)$.
Hence, $j_S^{t+1}(i)$ is equal to $j_B^t(i)$ if UE $i$ does handover, and it remains $j_S^{t}(i)$ otherwise.
As a result, the serving BSs for time slot $t+1$ are determined.

After $\mathrm{DM}_2^t$, the agent estimates the channel capacity between UEs and their backup BSs.
This is done during mini-time slot $B$ (see Fig. \ref{fig:fig1}).
Then, in order to find the amount of resources allocated to each UE, agent tries to find an optimal solution $\alpha^*(i)$ for $i\in\mathcal{U}$ which maximizes the sum of the achieved rate of all UEs in the zone while minimizing the amount number of UEs with average rate less than $\R_{\rm th}$. Formally, the agent finds the sub-optimal solution $\alpha^*(i)$ for $i\in\mathcal{U}$ for the following problem:
\begin{subequations}
\begin{alignat}{2}
&\max_{\{\alpha(i)\}_{i\in \mathcal{U}}}
&\;& \sum_{i\in\mathcal{U}} c^{t+1}(i,{j}^{t+1}_{S}(i)) \alpha(i) \nonumber\\
&&&\quad - \lambda_t 1\left\{ \frac{1}{K} \sum_{k=-1}^{K-2} \R^{t-k}(i)\leq \R_{\rm th}(i) \right\} \label{eq:optProb2} \\
&\mathrm{subject~to}
&& \sum_{i\in \mathcal{U}}\alpha(i) 1\{j_S^{t+1}(i) = j\} \leq 1,
\quad\forall j\in\mathcal{B}, \\
&&&\alpha(i)\geq 0,
\quad \forall i\in \mathcal{U}.
\end{alignat} \label{opt2}
\end{subequations}
This optimization is done in mini time slot $\mathrm{DM}_3^t$ (see Fig. \ref{fig:fig1}).
\begin{algorithm}[tp]
	\caption{Handover at the end of the time slot $t$. }\label{Alg2}
	\textbf{Inputs:} The set $\mathcal{U}_h$ includes all the UEs which require to do the handover.
	\begin{algorithmic}[1]
		\State{Choose $j^t_B(i) \forall i\in \mathcal{U}$, based on action $A(\mathcal{U}_h)$, where $j^t_B(i)=j^t_S(i)$ if $i\notin\mathcal{U}_h$.}
		\State{${j}^{t+1}_{S}(i) \gets {j}^{t}_{B}(i)$.}
		\State{During mini-time slot $B$,
			estimate channel from BS ${j}^{t}_{B}(i)$ toward UE $i$ in location $\ell^{t+1}(i)$ and calculate $c^{t+1}(i,{j}^{t}_{B}(i))$.}
		\State{Calculate required resource $\tilde{\alpha}(i), \forall i\in\mathcal{U}$ based on \eqref{eqn:tildealphai}.}
		\ForAll{$j\in\mathcal{B}$}
		\State{Evaluate $i_S^{t+1}(j)$ based on \eqref{eqn:iSt+1(j)} and find $N_S(j)$ from \eqref{eqn:lS(j)} .}
		\State{Find $\mathcal{I}^*_j$ from \eqref{eqn:I(j)} and  $\alpha^{t+1}(i)$ from \eqref{eqn:alpha(t+1)(i)} .}
		\EndFor
		\\{\textbf{Outputs:} $j^{t+1}_S(i)$ and $\alpha^{t+1}(i)$ for all $i\in\mathcal{U}$.}
	\end{algorithmic}
\end{algorithm}

Now, we explain a sub-optimal solution for \eqref{opt2}. As we mentioned before, we assume that $\lambda_t \gg 1$. Therefore, it is more important to minimize the number of UEs getting less than $\R_{\rm th}$ than to maximize the sum rate.
To this end, first, consider each BS separately.
If at most one UE is allocated to BS $j$, the BS allocates all of its resources to that UE, which is the best strategy for BS $j$.
Otherwise, find $\tilde{\alpha}(i)$, the amount of resources for each UE so that the rate is greater than the threshold:
\begin{equation} \label{eqn:tildealphai}
\tilde{\alpha}(i)
= \frac{\max\left\{0, K \R_{\rm th} - \sum_{k=0}^{k-2} \R^{t-k}(i)\right\}}{c^{t+1}(i,j_S^{t+1}(i))}.
\end{equation}
Define $i_S^{t+1}(j)$ as all UEs allocated to BS $j$:
\begin{equation} \label{eqn:iSt+1(j)}
i_S^{t+1}(j) := \{i \mid j_S^{t+1}(i) = j \}.
\end{equation}
Thus, $N_S(j)$, the maximum number of UEs which can be served by BS $j$ while satisfying the rate requirement, is
\begin{subequations}
\begin{alignat}{2}
N_S(j) =&\max_{\mathcal{I}_j\subseteq i_S^{t+1}(j)}
&\;& \lvert\mathcal{I}_j\rvert \\
&\mathrm{subject~to}
&& \sum_{i\in\mathcal{I}_j} \tilde{\alpha}(i) \leq 1.
\end{alignat} \label{eqn:lS(j)}
\end{subequations}
\!\!Next, for all $j\in\mathcal{B}$, by exhaustive search among all subsets of $i_S^{t+1}(j)$ with number of elements equal to $N_S(j)$, one finds $\mathcal{I}_j^*$ as
\begin{subequations}
\begin{alignat}{2}
\mathcal{I}_j^* = &\argmax_{\mathcal{I}_j\subseteq i_S^{t+1}(j)}
&\;& \sum_{i\in\mathcal{I}_j}\tilde{\alpha}(i) c^{t+1}(i,j) \\
&\mathrm{subject~to}
&& \lvert\mathcal{I}_j\rvert = N_S(j).
\end{alignat} \label{eqn:I(j)}
\end{subequations}
\!\!The remaining resources of BS $j$, which is $1\!-\!\sum_{i\in\mathcal{I}_j^*}\tilde{\alpha}(i)$, are allocated to the UE $\argmax_{i\in i_S^{t+1}(j)}c^{t+1}(i,j)$.
Hence,
\begin{IEEEeqnarray}{l}
\alpha^{t+1}(i) = \nonumber\\
\quad\begin{cases}
\tilde{\alpha}(i) & i\in\mathcal{I}_j^*, \\
1-\sum_{i\in\mathcal{I}_j^*}\tilde{\alpha}(i) & i=\argmax_{i\in i_S^{t+1}(j)}c^{t+1}(i,j), \\
0 & \mathrm{otherwise}.
\end{cases} \label{eqn:alpha(t+1)(i)} \IEEEeqnarraynumspace
\end{IEEEeqnarray}

\section{Numerical Results}
\label{results}
In this section, we evaluate the performance of our proposed handover method in the downlink of a mmWave network operating at 28 GHz with 4 BSs and 7 UEs. The mmWave links are generated as described in Section \ref{model_A}. All the UEs move through different trajectories with different directions in 100 seconds. We consider the pedestrian and the vehicular mobility models with speed $5$ km/h and $60$ km/h, respectively. We consider a zone of $100 \times 100 ~ m^2$ area. 
The main simulation parameters are listed in Table \ref{table1}.

The rate requirement are drawn from uniform distribution $\R_{\mathrm{th}}(i)\sim U[0.2,\R_{\mathrm{max}}]$ Gbps, $\forall i\in \mathcal{U}$, where $\R_{\mathrm{max}}$ can vary. The average rate per location is computed in $K=2$ previous time slots.
We define an episode of the learning process as a trajectory and use 5000 different realizations of the channels through the trajectories as the database of the learning process.

We compare the performance of our proposed method with two benchmarks. In the first benchmark, \texttt{Rand}, a backup solution is chosen randomly from one of the BSs in UE's vicinity. Although the \texttt{Rand} benchmark may be fast, it is not efficient and can be considered as a lower bound for the performance of our approach. The second benchmark, \texttt{WCS}, is inspired by the worst connection swapping algorithm in \cite{alizadeh2019load} which converges fast and provides a near optimal solution for user association problem. In this approach, the serving BSs of all the UEs needed to do a handover
are swapped with the ones maximizing the reward function. If the current connections are the best ones among all swapping possibilities (in \emph{switching} step), the serving BSs are swapped with another arbitrary UEs.

\begin{figure}[t]
	\centering
	\begin{subfigure}[t]{0.5\textwidth}
		{\footnotesize 
%
%
\definecolor{mycolor1}{rgb}{0.00000,0.44700,0.74100}%
\definecolor{mycolor2}{rgb}{0.85000,0.32500,0.09800}%
\definecolor{mycolor3}{rgb}{0.92900,0.69400,0.12500}%
\begin{tikzpicture}

\begin{axis}[%
width=0.85\columnwidth,
height=0.4\columnwidth,
at={(0\columnwidth,0\columnwidth)},
scale only axis,
xmin=-2.5,
xmax=52.5,
xlabel style={},
xlabel={Average number of the events where not meet the rate requirement},
ymin=0,
ymax=0.4,
ylabel style={},
ylabel={Probability density function},
axis background/.style={fill=white},
legend style={legend cell align=left, align=left, draw=white!15!black}
]
\addplot[ybar, bar width=0.35, fill=green, area legend] table[row sep=crcr] {%
0	0.13\\
1	0.0714285714285714\\
2	0.0642857142857143\\
3	0\\
4	0.0514285714285714\\
5	0.05\\
6	0.0371428571428571\\
7	0\\
8	0.0414285714285714\\
9	0.0428571428571429\\
10	0.0342857142857143\\
11	0\\
12	0.0271428571428571\\
13	0.0314285714285714\\
14	0.01714285714285714\\
15	0\\
16	0.06\\
17	0.0114285714285714\\
18	0.00571428571428571\\
19	0\\
20	0.0157142857142857\\
21	0.02428571428571429\\
22	0.00571428571428571\\
23	0\\
24	0.0114285714285714\\
25	0.04142857142857143\\
26	0.00428571428571429\\
27	0\\
28	0.00285714285714286\\
29	0.00285714285714286\\
30	0.00142857142857143\\
31	0\\
32	0\\
33	0\\
34	0.00285714285714286\\
35	0\\
36	0.0314285714285714\\
37	0.00314285714285714\\
38	0.00414285714285714\\
39	0\\
40	0.0157142857142857\\
41	0.0128571428571429\\
42	0.0185714285714286\\
43	0\\
44	0.00714285714285714\\
45	0.00714285714285714\\
46	0.00571428571428571\\
47	0\\
48	0.00000005714285714\\
49	0.03\\
50	0.000002857142857143\\
51	0.00285714285714286\\
};
\addplot[forget plot, color=blue] table[row sep=crcr] {%
0.511111111111111	0\\
51.4888888888889	0\\
};
\addlegendentry{Our approach}

\addplot[ybar, bar width=0.2, fill=blue, area legend] table[row sep=crcr] {%
0.8	0.00285714285714286\\
1.8	0.00285714285714286\\
2.8	0.01\\
3.8	0.00857142857142857\\
4.8	0.0214285714285714\\
5.8	0.0214285714285714\\
6.8	0.0214285714285714\\
7.8	0.0142857142857143\\
8.8	0.0257142857142857\\
9.8	0.0342857142857143\\
10.8	0.0314285714285714\\
11.8	0.02\\
12.8	0.0428571428571429\\
13.8	0.0185714285714286\\
14.8	0.0342857142857143\\
15.8	0.0357142857142857\\
16.8	0.0242857142857143\\
17.8	0.0371428571428571\\
18.8	0.0371428571428571\\
19.8	0.0385714285714286\\
20.8	0.0442857142857143\\
21.8	0.0514285714285714\\
22.8	0.0357142857142857\\
23.8	0.0314285714285714\\
24.8	0.0328571428571429\\
25.8	0.0371428571428571\\
26.8	0.04\\
27.8	0.0414285714285714\\
28.8	0.0342857142857143\\
29.8	0.0342857142857143\\
30.8	0.0314285714285714\\
31.8	0.0214285714285714\\
32.8	0.0171428571428571\\
33.8	0.01\\
34.8	0.0128571428571429\\
35.8	0.00857142857142857\\
36.8	0.00571428571428571\\
37.8	0.00857142857142857\\
38.8	0.00857142857142857\\
39.8	0.000285714285714286\\
40.8	0.000285714285714286\\
41.8	0.00285714285714286\\
42.8	0\\
43.8	0\\
44.8	0.00142857142857143\\
45.8	0\\
46.8	0.00110001\\
47.8	0\\
48.8	0.0123\\
49.8	0\\
50.8	0\\
};
\addplot[forget plot, color=white!15!black] table[row sep=crcr] {%
0.511111111111111	0\\
51.4888888888889	0\\
};
\addlegendentry{Rand}

\addplot[ybar, bar width=0.2, fill=red, area legend] table[row sep=crcr] {%
1.2	0.182857142857143\\
2.2	0.0357142857142857\\
3.2	0.02\\
4.2	0.00428571428571429\\
5.2	0.00428571428571429\\
6.2	0.00428571428571429\\
7.2	0.01\\
8.2	0.00571428571428571\\
9.2	0.00428571428571429\\
10.2	0.00428571428571429\\
11.2	0.0114285714285714\\
12.2	0.00142857142857143\\
13.2	0.00714285714285714\\
14.2	0.07\\
15.2	0.00857142857142857\\
16.2	0.00571428571428571\\
17.2	0.00142857142857143\\
18.2	0.00428571428571429\\
19.2	0.00857142857142857\\
20.2	0.00714285714285714\\
21.2	0.0114285714285714\\
22.2	0.00857142857142857\\
23.2	0.00714285714285714\\
24.2	0.00285714285714286\\
25.2	0.00714285714285714\\
26.2	0.0114285714285714\\
27.2	0.00428571428571429\\
28.2	0.00857142857142857\\
29.2	0.00571428571428571\\
30.2	0.00571428571428571\\
31.2	0.0514285714285714\\
32.2	0.0171428571428571\\
33.2	0.0214285714285714\\
34.2	0.0242857142857143\\
35.2	0.0214285714285714\\
36.2	0.0514285714285714\\
37.2	0.0142857142857143\\
38.2	0.0142857142857143\\
39.2	0.0114285714285714\\
40.2	0.0171428571428571\\
41.2	0.00714285714285714\\
42.2	0.00142857142857143\\
43.2	0\\
44.2	0.00142857142857143\\
45.2	0\\
46.2	0.00142857142857143\\
47.2	0\\
48.2	0\\
49.2	0\\
50.2	0.155714285714286\\
51.2	0.00571428571428571\\
};
\addplot[forget plot, color=white!15!black] table[row sep=crcr] {%
0.511111111111111	0\\
51.4888888888889	0\\
};
\addlegendentry{WCS}

\end{axis}

\end{tikzpicture}%
}
		\caption{}
		\label{fig2:(a)}		
	\end{subfigure}
	\begin{subfigure}[t]{0.5\textwidth}
		{\footnotesize 
%
%
\begin{tikzpicture}
\begin{axis}[%
width=0.85\columnwidth,
height=0.4\columnwidth,
at={(0\columnwidth,0\columnwidth)},
scale only axis,
xmin=-2.4,
xmax=50.4,
xlabel style={},
xlabel={Average number of handovers},
ymin=0,
ymax=0.7,
ylabel style={},
ylabel={Probability density function},
axis background/.style={fill=white},
legend style={legend cell align=left, align=left, draw=white!15!black}
]
\addplot[ybar, bar width=0.32, fill=green, draw=black, area legend] table[row sep=crcr] {%
1	0.374285714285714\\
2	0.625714285714286\\
3	0\\
4	0\\
5	0\\
6	0\\
7	0\\
8	0\\
9	0\\
10	0\\
11	0\\
12	0\\
13	0\\
14	0\\
15	0\\
16	0\\
17	0\\
18	0\\
19	0\\
20	0\\
21	0\\
22	0\\
23	0\\
24	0\\
25	0\\
26	0\\
27	0\\
28	0\\
29	0\\
30	0\\
31	0\\
32	0\\
33	0\\
34	0\\
35	0\\
36	0\\
37	0\\
38	0\\
39	0\\
40	0\\
41	0\\
42	0\\
43	0\\
44	0\\
45	0\\
46	0\\
47	0\\
48	0\\
49	0\\
};
\addplot[forget plot, color=white!15!black] table[row sep=crcr] {%
0.511111111111111	0\\
49.4888888888889	0\\
};
\addlegendentry{Oue approach}

\addplot[ybar, bar width=0.2, fill=blue, draw=black, area legend] table[row sep=crcr] {%
0.8	0.00142857142857143\\
1.8	0.00428571428571429\\
2.8	0.00714285714285714\\
3.8	0.00857142857142857\\
4.8	0.0128571428571429\\
5.8	0.02\\
6.8	0.0214285714285714\\
7.8	0.0228571428571429\\
8.8	0.0471428571428571\\
9.8	0.0628571428571429\\
10.8	0.0542857142857143\\
11.8 0.06\\
12.8	0.06\\
13.8	0.0642857142857143\\
14.8	0.0828571428571429\\
15.8	0.07\\
16.8	0.0614285714285714\\
17.8	0.0585714285714286\\
18.8	0.0671428571428571\\
19.8	0.0514285714285714\\
20.8	0.0457142857142857\\
21.8	0.0328571428571429\\
22.8	0.0214285714285714\\
23.8	0.0114285714285714\\
24.8	0.0142857142857143\\
25.8	0.0114285714285714\\
26.8	0.00714285714285714\\
27.8	0.00857142857142857\\
28.8	0.00285714285714286\\
29.8	0.00142857142857143\\
30.8	0.00285714285714286\\
31.8	0.00142857142857143\\
32.8	0\\
33.8	0\\
34.8	0\\
35.8	0\\
37	0\\
38	0\\
39	0\\
40	0\\
41	0\\
42	0\\
43	0\\
44	0\\
45	0\\
46	0\\
47	0\\
48	0\\
49	0\\
};
\addplot[forget plot, color=white!15!black] table[row sep=crcr] {%
0.511111111111111	0\\
49.4888888888889	0\\
};
\addlegendentry{Rand}

\addplot[ybar, bar width=0.2, fill=red, draw=black, area legend] table[row sep=crcr] {%
1.2	0.321428571428571\\
2.2	0.0885714285714286\\
3.2	0.0657142857142857\\
4.2	0.00714285714285714\\
5.2	0.0128571428571429\\
6.2	0.00285714285714286\\
7.2	0.0114285714285714\\
8.2	0.00285714285714286\\
9.2	0.00142857142857143\\
10.2	0.00714285714285714\\
11.2	0.00285714285714286\\
12.2	0.00714285714285714\\
13.2	0.0114285714285714\\
14.2	0.00285714285714286\\
15.2	0.00285714285714286\\
16.2	0.00857142857142857\\
17.2	0.00428571428571429\\
18.2	0.01\\
19.2	0.01\\
20.2	0.00285714285714286\\
21.2	0.0128571428571429\\
22.2	0.01\\
23.2	0.00714285714285714\\
24.2	0.0171428571428571\\
25.2	0.00571428571428571\\
26.2	0.0157142857142857\\
27.2	0.0114285714285714\\
28.2	0.02\\
29.2	0.0228571428571429\\
30.2	0.0228571428571429\\
31.2	0.0257142857142857\\
32.2	0.0371428571428571\\
33.2	0.0228571428571429\\
34.2	0.0285714285714286\\
35.2	0.0314285714285714\\
36.2	0.0314285714285714\\
37.2	0.0257142857142857\\
38.2	0.0185714285714286\\
39.2	0.0157142857142857\\
40.2	0.00714285714285714\\
41.2	0.00714285714285714\\
42.2	0\\
43.2	0.00571428571428571\\
44.2	0.00285714285714286\\
45.2	0\\
46.2	0\\
47.2	0.00571428571428571\\
48.2	0.00285714285714286\\
49.2	0.00142857142857143\\
};
\addplot[forget plot, color=white!15!black] table[row sep=crcr] {%
0.511111111111111	0\\
49.4888888888889	0\\
};
\addlegendentry{WCS}

\end{axis}
\end{tikzpicture}%
}
		\caption{}
		\label{fig2:(b)}	
	\end{subfigure}
	\caption{(a) Average number of the UEs with rates lower than the minimum rate requirements, and (b) average number of handovers.}
	\label{fig2:2}
\end{figure}
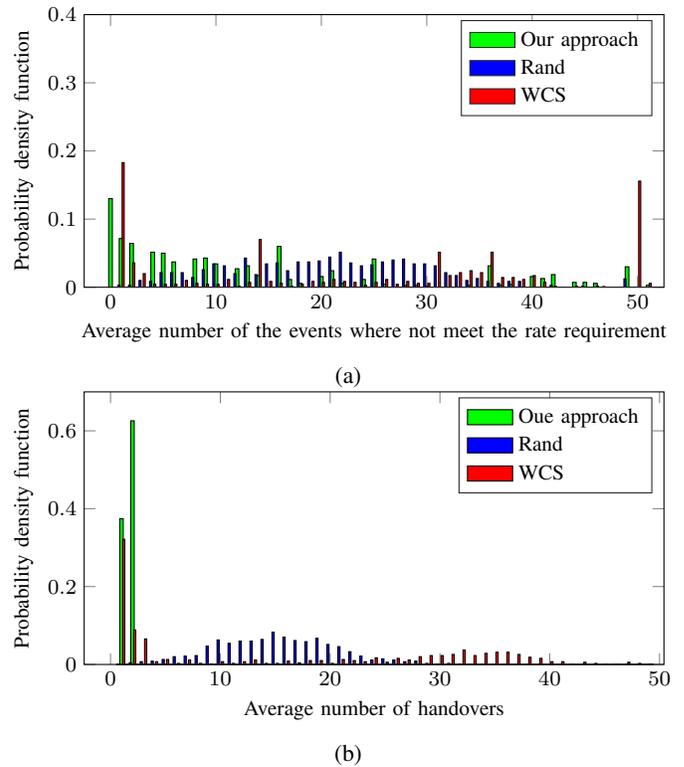

Table \ref{table2} and Fig. \ref{fig2:2} illustrate three objectives ($F_1$, $F_2$, and $F_3$) of optimization problem \eqref{optimization} for all algorithms. Compared to the benchmarks, our approach minimizes the number of handovers and the outage events (average number of events where not meet the rate requirement) while maximizes the sum rate of all the UEs. It happens because our handover algorithm prioritizes reducing the number of outage events and the number of handovers over maximizing the sum rate. As a result our algorithm provides more reliable connection.



\section{Conclusions}
\label{conclusions}
In this paper, we investigated the problem of the load balancing handover in multi-user mobile mmWave networks. We leverage machine-learning tools to approximate the non-convex association problem. Numerical evaluation confirmed that our algorithm maximizes the sum rate of all the UEs while guaranteeing a minimum rate requirement, as well as preventing frequent handovers for all the UEs in comparison with the benchmarks.
\begin{table}[t]
	\begin{center}
		\caption{Simulation parameters. }
		\label{table1}
		\begin{tabular}{|c|c|} 
			\hline
			\textbf{Parameters} & \textbf{Values in Simulations} \\
			\hline
			\hline
			$\BS$ transmit power & 30 dBm\\
			Thermal noise power	&$\sigma^2$=-174 dBm/Hz\\
			Signal bandwidth & $W$=500 MHz \\
			BS antenna & $8\times8$\\
			UE antenna & $4\times4$\\
			
			LoS path loss exponent & $\hat{n}_{\LoS}=3$ \\
			NLoS path loss exponent&$\hat{n}_{\NLoS}=4$ \\

			Discount factor&	$\gamma= 0.99$ \\
			Mini-batch size& $M=32$\\
			$\lambda_t$& 20\\
			$\lambda_h$&10\\
			\hline
		\end{tabular}
	\end{center}
\end{table}
\begin{table}[t]
	\begin{center}
		\caption{Average sum rate of all the UEs.}
		\label{table2}
		\begin{tabular}{|c|c|} 
			\hline
			\textbf{Methods} & \textbf{Average sum rate of all the UEs (Gbps)} \\
			\hline
			\hline
			Our approach & 190.1486\\
			\hline
			Benchmark 1	& 188.7487\\
			\hline
			Benchmark 2	&191.2312\\
			\hline
		\end{tabular}
	\end{center}
\end{table}

\bibliographystyle{IEEEtran}
\bibliography{IEEEabrv,ref_bib2}
\end{document}